\documentclass[twocolumn,pra,showpacs,superscriptaddress]{revtex4}

\usepackage{graphicx}
\usepackage{dcolumn}
\usepackage{bm}
\usepackage{amssymb}
\usepackage{amsmath}
\usepackage{extarrows}

\begin{document}

%\title{Simulation and detection of three-dimensional Weyl semimetal physics in a two-dimensional optical lattice}

\title{Simulating and exploring Weyl semimetal physics with cold atoms in a two-dimensional optical lattice}

\author{Dan-Wei Zhang}
\email{zdanwei@hku.hk}
\affiliation{Department of Physics and Center of Theoretical and Computational Physics, The University of Hong Kong,
Pokfulam Road, Hong Kong, China}

\author{Shi-Liang Zhu}
\email{slzhu@nju.edu.cn} \affiliation{National Laboratory of Solid
State Microstructures and School of Physics, Nanjing University,
Nanjing 210093, China} \affiliation{Synergetic Innovation Center of Quantum Information and Quantum
Physics, University of Science and Technology of China, Hefei
230026, China}

\author{Z. D. Wang}
\email{zwang@hku.hk} \affiliation{Department of Physics and Center of Theoretical and Computational Physics, The University of Hong Kong, Pokfulam Road, Hong Kong, China}

\begin{abstract}
We propose a scheme to simulate and explore Weyl semimetal physics with
ultracold fermionic atoms in a two-dimensional square optical
lattice subjected to experimentally realizable spin-orbit coupling
and an artificial dimension from an external parameter space,
which may increase experimental feasibility compared with the cases in three-dimensional optical lattices.
It is shown that this system with a tight-binding model is able to describe essentially
three-dimensional Weyl semimetals with tunable Weyl points. The relevant topological properties are also addressed by means of the Chern number and
the gapless edge states. Furthermore, we illustrate that the
mimicked Weyl points can be experimentally detected by measuring
the atomic transfer fractions in a Bloch-Zener oscillation, and the
characteristic topological invariant can be measured with the
particle pumping approach.
\end{abstract}

\date{\today}


\maketitle

\section{introduction}

Weyl semimetal (WSM), as an exotic topologically nontrivial state of matter
in three dimensions (3D) \cite{Kane,Qi,Wan,Balents,Burkov,Xu,Delplace,Zhao}, has a singly-degenerate band structure with paired
bulk band crossings, named as Weyl points. A linear dispersion
relation in all three momentum directions exhibits near a Weyl
point, with low-energy excitations resembling the well-known Weyl
fermions in particle physics. Each Weyl point has a topological
charge, namely the Chern number on the gapped 2D sphere enclosing
a Weyl point in momentum space, and a pair of points with opposite charges support topologically
protected gapless Fermi arc surface states
\cite{Wan,Balents,Burkov,Xu,Delplace,Zhao,Volovik,Zhao2013}. The Weyl points and
the Fermi arcs in a WSM are fundamentally interesting,
and are expected to give rise to exotic phenomena absent in fully
gapped topological phases, such as anomalous electromagnetic
responses \cite{Zhao,Zyuzin,Parameswaran}. Recently, more and more
efforts have been made on experimental realization of WSMs in real materials \cite{Xu1,Huan,Lv,Xu2,Zhang} and in
artificial systems, such as analogous Weyl points in photonic
crystals \cite{Lu1,Lu2}.

On the other hand, ultracold atomic gases in optical lattices
provide a powerful platform to simulate various quantum states of
matter \cite{Lewenstein,ZhangFP}. In particular, recent
experimental advances in engineering spin-orbit coupling and
artificial gauge field for ultracold atoms
\cite{Lin1,Lin2,ZhangJ,Cheuk,Dalibard} have pushed this system to
the forefront for studying topological states of matter
\cite{Shao,Goldman2010,Beri,Alba,Zhu,Sun,Deng,Price,Liu2013,Demler,Bermudez,Wang,Bloch2013,Jotzu,Bloch2015}.
Another recently developed technique related to the realization of
topological phases in cold atom systems (also in photonic
quasicrystals, e.g. see Ref. \cite{quasicrystals}) is extending an
artificial dimension in optical lattices provided by an external
cyclical parameter \cite{Chen,Mei} or the internal atomic degrees
of freedom \cite{Celi,Spielman}. Using this technique, one can
study the physics of topological states in optical lattices
attributed to dimensions higher than their own. Nowadays, the Zak
phase in topologically nontrivial Bloch bands realized in 1D
optical lattices has been measured \cite{Bloch2013}. The 2D Chern
insulators and the Hofstadter bands have been also realized
experimentally in 2D optical lattices with the Chern number
therein being successfully probed \cite{Jotzu,Bloch2015}. The
experimental observation of chiral edge states in 1D
optical lattices subjected a synthetic magnetic field and an
artificial dimension has been reported \cite{Spielman}.
By stacking multilayers of 2D atomic Chern or Hofstadter insulators, it was proposed to realize
WSMs in 3D optical lattices \cite{Jiang,Buljan,He}.
This construction method was extended to simulate WSMs
in 1D double-well optical lattices with two degrees of freedoms modulating the hopping terms \cite{Ganeshan}, but it seems
extremely hard to independently control them in practical experiments.
Due to the elusive nature of WSMs and their intrinsic
exotic properties, other feasible schemes for their experimental realization
or simulation would be still of great value. In particular, the simulation and
detection of the Weyl points and the relevant topological
properties of WSMs with an artificial dimension in feasible 2D cold atomic systems are still badly awaited.

In this paper, we propose a scheme to simulate and explore essential physics of WSMs
with ultracold fermionic atoms in a
2D square optical lattice, subjected to experimentally realizable
spin-orbit coupling and an artificial dimension from an external
cyclical parameter. We first show that this system with a
tight-binding model is able to mimic 3D WSMs
with tunable Weyl points. The topological properties in this
system are then further investigated by calculating the out-of-plane
momentum, say $k_z$, -dependent Chern number and the gapless edge
states. Finally, we propose practical methods for the experimental
detection of the mimicked Weyl points and the characteristic
topological invariant in the proposed cold atomic system. With
numerical simulations, we demonstrate that the analogous Weyl
points in this system can be sharply probed by measuring the
Landau-Zener tunneling to the excited band after a Bloch
oscillation and the $k_z$-dependent Chern number can directly be
extracted from the center shift of the hybrid Wannier functions,
both of which are measurable with cold atoms in optical lattices
by the time-of-flight images. Comparing with the schemes evolving complex synthetic magnetic fluxes in 3D optical lattices \cite{Jiang,Buljan,He}, our scenario not only is different but also may increase experimental feasibility. In addition, the proposed system with high tunability and convenient detection ways may provide a promising platform for exploring exotic WSM physics.

The paper is organized as follows. Section II
introduces the cold atom system with a tight-binding model for
simulating the WSM phase with tunable Weyl points.
In Sec. III, we elaborate the topological properties in this model
by calculating the $k_z$-dependent Chern number and gapless edge
states. In section IV, we propose practical methods for the
experimental detection of the mimicked Weyl points and the
characteristic topological invariant in the proposed cold atomic
system with numerical simulations. Finally, a short conclusion is
given in Sec. V.

\section{model}

In this section, we construct a tight-binding model in a 2D square
lattice subjected to an artificial  dimension for simulating the
tunable WSM phase. Let us consider a non-interacting
spin-1/2 ultracold degenerate fermionic gas (labeled as spins
$\uparrow$ and $\downarrow$) loaded in a 2D square optical lattice
in the $xz$ plane. In the tight-binding regime, the atomic hopping
between two nearest-neighbor lattice sites can be spin-conserved
hopping or spin-flip hopping which can be achieved by Raman
coupling between the two spin states $|\uparrow\rangle$ and
$|\downarrow\rangle$ \cite{Dalibard}. The tight-binding
Hamiltonian of this cold atom system is considered to be
\begin{eqnarray}
\mathcal{\hat{H}}=\hat{H}_L+\hat{H}_{SOC}+\hat{H}_P,
\end{eqnarray}
which consists of the spin-conserved hopping in the 2D lattice
$\hat{H}_L$, the spin-flip hopping along the $x$  axis
$\hat{H}_{SOC}$, and $\hat{H}_P$ representing additional external
coupling potential:
\begin{eqnarray}
&&\hat{H}_L = -\sum_{\mathbf{i},\sigma,\eta}t_{\sigma}\left(\hat{a}^{\dag}_{\mathbf{i},\sigma}\hat{a}_{\mathbf{i}+\hat{\eta},\sigma}+\text{h.c.}\right),\\
&&\hat{H}_{SOC} =-t_s\sum_{\mathbf{i}}\left(\hat{a}^{\dag}_{\mathbf{i},\uparrow}\hat{a}_{\mathbf{i}+\hat{x},\downarrow}-\hat{a}^{\dag}_{\mathbf{i},\uparrow}\hat{a}_{\mathbf{i}-\hat{x},\downarrow}+\text{h.c.}\right),\\
&&\hat{H}_P =\Gamma_z\sum_{\mathbf{i}}\left(\hat{a}^{\dag}_{\mathbf{i},\uparrow}\hat{a}_{\mathbf{i},\uparrow}-\hat{a}^{\dag}_{\mathbf{i},\downarrow}\hat{a}_{\mathbf{i},\downarrow}\right) \\\nonumber&&~~~~~~~+\Gamma_x\sum_{\mathbf{i}}\left(\hat{a}^{\dag}_{\mathbf{i},\uparrow}\hat{a}_{\mathbf{i},\downarrow}+\hat{a}^{\dag}_{\mathbf{i},\downarrow}\hat{a}_{\mathbf{i},\uparrow}\right).
\end{eqnarray}
Here $\hat{a}_{\mathbf{i},\sigma}$ ($\hat{a}^{\dag}_{\mathbf{i},\sigma}$) annihilates (creates) a fermion on site $\mathbf{i}=(i_x,i_z)$ with spin $\sigma=\{\uparrow,\downarrow\}$ and $\eta=\{x,z\}$. The spin-dependent hopping amplitude $t_{\sigma}$ in $\hat{H}_L$ is assumed to be $t_{\uparrow}=-t_{\downarrow}=t_0$, which can be achieved in experiments by Raman laser or modulation engineering hopping \cite{Dalibard}, such as laser-induced Peierls phase factor $e^{i\pi}$ for the $|\downarrow\rangle$ atomic hopping. The spin-flip hopping term $\hat{H}_{SOC}$ can be induced by the experimentally realized spin-orbit coupling with equal Rashba and Dresselhaus amplitudes through a two-photon Raman process \cite{Lin2,ZhangJ,Cheuk,Dalibard}. Finally, the Hamiltonian $\hat{H}_P$ is composed of a Zeeman potential and an additional coupling term on the spin states. The Zeeman potential is generated by a external magnetic field and the coupling term is induced by additional Raman lasers \cite{Gross}. The Zeeman and Raman terms can be turned fully and independently in cold atom experiments with long coherence time \cite{Gross}, and thus we choose the parameters in the following forms
\begin{eqnarray}
\Gamma_z = m_z-2t_0\cos\theta,~~\Gamma_x = 2t_s\sin\theta,
\end{eqnarray}
where $m_z$ is a constant parameter and $\theta$ is a cyclical parameter that can vary from $\theta=-\pi$ to $\theta=\pi$.

Under the period boundary condition, the total Hamiltonian of the system can be rewritten as
\begin{eqnarray}
\mathcal{\hat{H}} = \sum_{{\boldsymbol k},\sigma\sigma'}\hat{a}^{\dag}_{{\boldsymbol k}\sigma}\left[H({\boldsymbol k})\right]_{\sigma\sigma'}\hat{a}_{{\boldsymbol k}\sigma'},
\end{eqnarray}
where $\hat{a}_{{\boldsymbol k}\sigma}=1/\sqrt{V}\sum_{\boldsymbol k}e^{-i{\boldsymbol k}\cdot\mathbf{i}}\hat{a}_{\mathbf{i},\uparrow}$ is the annihilation operators in momentum space ${\boldsymbol k}=(k_x,k_z)$, and $H({\boldsymbol k})$ is the Bloch Hamiltonian given by $H({\boldsymbol k})= \vec{d}({\boldsymbol k})\cdot\vec{\sigma}$. Here $\vec{\sigma}=(\sigma_x,\sigma_y,\sigma_z)$ are the Pauli matrices acting on the atomic components and $\vec{d}=(d_x,d_y,d_z)$ is the Bolch vectors: $d_x = 2t_s\sin\theta$, $d_y = 2t_s\sin k_x$, and $d_z = m_z-2t_0\cos k_z-2t_0\cos k_x-2t_0\cos \theta$, with lattice spacing $a\equiv1$ and $\hbar\equiv1$ hereafter. The energy spectrum of the system is then given by $E_{\pm}({\boldsymbol k},\theta)= \pm |\vec{d}({\boldsymbol k},\theta)|$.
%%
%\begin{eqnarray}
%E_{\pm}({\boldsymbol k},\theta)= \pm \sqrt{d_x^2+d_y^2+d_z^2}
%\end{eqnarray}
%%
Treating the cyclical parameter $\theta$ as the Bloch momentum $k_y$, which acts as an artificial dimension of the $y$ axis under periodical condition along this direction, we can consider the 2D system in a 3D parameter space with the 3D-extended momentum $\boldsymbol{\tilde{k}}=(k_x,\theta,k_z)$. The bulk energy bands are fully gapped except the points that satisfy the following conditions:
\begin{eqnarray}
 &&\sin k_x = \sin \theta =0,\\
 && m_z-2t_0(\cos k_z+\cos k_x+\cos \theta)=0,
\end{eqnarray}
which determine the boundaries between the mimicked WSM phase and the bulk insulating phases as discussed in the following.

We now show that the WSM phase can be simulated in this
system. Without loss of generality, we focus on the  parameters
$t_0>0$, $t_s>0$ and $m_z\geq0$ in the system in the rest of the
this paper. The first condition of Eq. (7) requires
$k_x=\{0,\pi\}$ and $\theta=\{0,\pi\}$, and consequently the
second condition of Eq. (8) reduces to:
\begin{align}
&m_z-2t_0\cos k_z=0,~\text{for}~(k_x,\theta)=(0,\pi)~\text{and}~(\pi,0);\\
&m_z-2t_0\cos k_z-4t_0=0,~\text{for}~(k_x,\theta)=(0,0);\\
&m_z-2t_0\cos k_z+4t_0=0,~\text{for}~(k_x,\theta)=(\pi,\pi).
\end{align}
When one of the three equations (9-11) is fulfilled with the
double-valued $k_z\in[-\pi,\pi]$, a pair of degenerate points
will exhibit in extended 3D band structure, which can be regarded
as analogous Weyl points in the $\boldsymbol{\tilde{k}}$ space.
For the considered parameter regime, Eq. (11) can not be
fulfilled, while Eq. (9) gives rise to two pairs of analogous Weyl
points denoted by
$\boldsymbol{W}_{1,\pm}=(0,\pi,\pm\arccos\frac{m_z}{2t_0})$ and
$\boldsymbol{W}_{2,\pm}=(\pi,0,\pm\arccos\frac{m_z}{2t_0})$ for
$m_z\leqslant2t_0$. Eq. (10) gives rise to a single pair of Weyl
points
$\boldsymbol{W}_{\pm}=(0,0,\pm\arccos\frac{m_z-4t_0}{2t_0})$ for
$2t_0\leqslant m_z\leqslant6t_0$. In Fig. 1, we plot the extended
3D bulk band structure $E_+(k_x,\theta,k_z)$ for the typical
parameters $m_z$ with fixed hopping energy $t_0=t_s=1$. Therefore,
the mimicked Weyl points can be found in the positions where
$E_+=0$. When $m_z=1$ [Fig. 1(a)], two pairs of analogous Weyl points locate
at $(k_x,\theta,k_z)=(0,\pi,\pm\pi/3)$ and $(\pi,0,\pm\pi/3)$ as
expected, and the critical case appears when $m_z=2$ [Fig. 1(b)],
with the two pairs merging at $(0,\pi,0)$ and $(\pi,0,0)$, and
another pair being created at $(0,0,\pm\pi)$. When $m_z=4$ [Fig.
1(c)], the newly created pair of analogous Weyl points move to the position
$(0,0,\pm\pi/2)$, and then merge at the location $(0,0,0)$ when
increasing the parameter to the critical value $m_z=6$. From Fig.
1, we can find that the number and location of the analogous Weyl
points are able to be tuned just by varying the parameter $m_z$.
This demonstrates that one can conveniently create, move, and
merge the mimicked Weyl points in this cold atom system.

\begin{figure}[tbph]
\centering
\includegraphics[width=8.5cm]{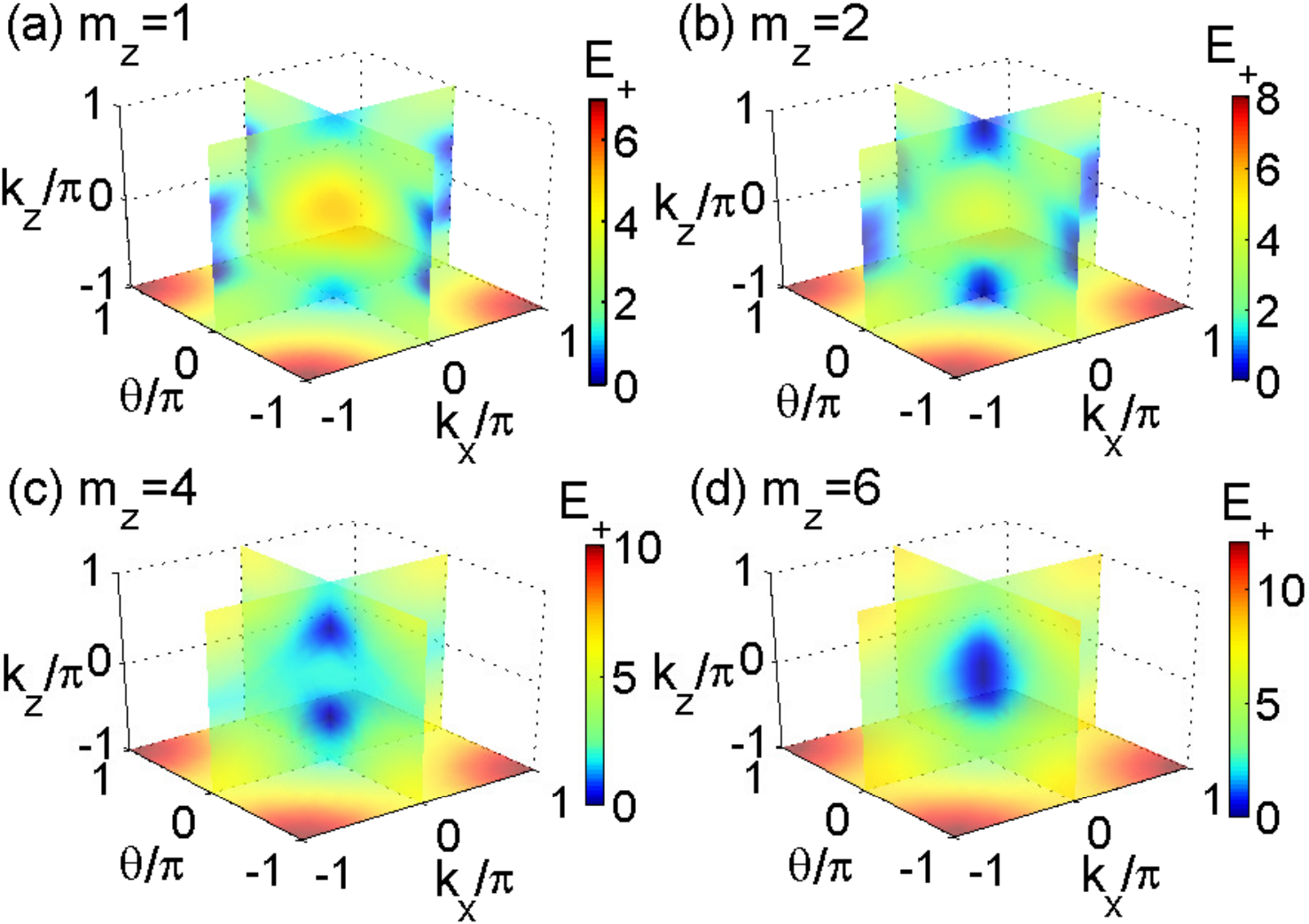}
\caption{(Color online) Color-coded extended 3D bulk band structure $E_+(k_x,\theta,k_z)$ for the typical parameters $m_z$.
(a) $m_z=1$ with two pairs of analogous Weyl points located at $(k_x,\theta,k_z)=(0,\pi,\pm\pi/3)$ and $(\pi,0,\pm\pi/3)$; (b) $m_z=2$, critical case with two pairs of analogous Weyl points merging at $(0,\pi,0)$ and $(\pi,0,0)$, and another pair appear at $(0,0,\pm\pi)$; (c) $m_z=4$ with one pair of analogous Weyl points located at $(0,0,\pm\pi/2)$ ; and (d) $m_z=6$, critical case with the pair of analogous Weyl points merging at $(0,0,0)$. Other parameters in (a-d) are $t_0=1$ as the energy unit and $t_s=1$.}
\end{figure}

Without loss of generality, hereafter we focus on the simplest
case of a singe pair of Weyl points $\boldsymbol{W}_{\pm}$,  where
the bulk bands only touch at the two distinct points in the
$\boldsymbol{\tilde{k}}$ space. We can expand the Bloch
Hamiltonian around the two mimicked Weyl points up to liner order
in $\boldsymbol{\tilde{k}}$, and then obtain the low-energy
effective Hamiltonian
\begin{eqnarray}
&&H_{\mathrm{W},\pm} = v_xq_x\sigma_y+v_yq_y\sigma_x\pm v_zq_z\sigma_z,
\end{eqnarray}
where $v_x=v_y=2t_s$ and $v_z=2t_0$ are the effective Fermi velocity, and $\boldsymbol{q}\equiv(q_x,q_y,q_z)=\boldsymbol{\tilde{k}}-\boldsymbol{W}_{\pm}$ for the two points, respectively. The Hamiltonian (12) is an analogous anisotropic Weyl Hamiltonian for the Weyl fermions and can be written as $H_{\mathrm{W},\pm} = \sum_{i,j}q_i\alpha_{ij}\sigma_j$, where $[\alpha_{ij}]$ is a $3\times3$ matrix with elements $\alpha_{xy}=\alpha_{yx}=2t_s$, $\alpha_{zz}=\pm2t_0$ and zero otherwise. Thus the chirality of the two Weyl points $\boldsymbol{W}_{\pm}$ can be defined as $\chi_{\pm}=\mathrm{sign}(\det[\alpha_{ij}])=\pm1$, respectively.

\section{Topological properties}

The opposite chirality of the paired analogous Weyl points
reflects the topological properties of this system, which can be characterized by the corresponding topological charge. Actually, a mimicked Weyl point here can be regarded as a monopole in the
$\boldsymbol{\tilde{k}}$ space, whose topological charge equals its
chirality \cite{Volovik,Zhao2013}. The paired Weyl points are robust to perturbations
which add a $\sigma_{x,y,z}$ term to the Hamiltonian because
their charge is quantized: the monopole charge cannot vary under a
continuous change of the Hamiltonian and they can only change at
creation or annihilation of a monopole-antimonopole pair. Moreover, it was shown recently that the gapless edge modes of a WSM are also robust against disorders \cite{Zhao}.

To further look into the topological properties of this system, we
consider the Hamiltonian (1) with the so-called dimension
reduction method. We treat $k_z$ as an effective parameter and
reduce the original system to a ($k_z$-modified) collection of effective 2D systems, as tight-binding chains along the $x$
axis. Such a reduction method is valid for the bulk system and the
reduced chains if $k_z$ is a good quantum number. For a fixed
$k_z$, the reduced Bloch Hamiltonian $H(k_x,\theta,k_z)\rightarrow
H_{k_z}(k_x,\theta)$, with
\begin{eqnarray}
&&H_{k_z}(k_x,\theta)=2t_s\sin\theta\sigma_x+2t_s\sin k_x\sigma_y\\ \nonumber
&&~~~~~~~~~~~~~~~~+(M_z-2t_0\cos k_x-2t_0\cos\theta)\sigma_z
\end{eqnarray}
acting as an effective 2D band structure in the $k_x-\theta$
parameter space, where $M_z=m_z-2t_0\cos k_z$. Since the extended
3D bulk bands are fully gapped when $k_z\neq\pm k_z^c$ with
$k_z^c=\arccos[(m_z-4t_0)/2t_0]$, in this case,
$H_{k_z}(k_x,\theta)$ describes a system of the effective 2D Chern
insulator. Therefore, one can have a well-defined Chern number
\begin{eqnarray}
\nonumber&&C_{k_z}=\frac{1}{4\pi} \int_{-\pi}^{\pi}dk_x\int_{-\pi}^{\pi}d\theta~\hat{d}\cdot\left(\partial_{k_x}\hat{d}\times\partial_{\theta}\hat{d}\right)\\
&&~~~~~= \left\{
                                  \begin{array}{ll}
                                    -\text{sign}(M_z), ~~& 0<|M_z|<4t_0; \\
                                     0, ~~& |M_z|>4t_0.
                                  \end{array}
                                \right.
\end{eqnarray}
where $\hat{d}\equiv \vec{d}/|\vec{d}|$. For our case within the parameter regime $2t_0\leqslant m_z\leqslant6t_0$, we can obtain that $C_{k_z}=-1$ when $k_z\in(-k_z^c,k_z^c)$ and otherwise $C_{k_z}=0$. For each $k_z$-fixed chain with a
nonzero Chern number, according to the bulk-edge correspondence and the index theorem \cite{Zhao2014}, there must be topologically protected gapless if an edge is created, i.e. finite lattice sites along the $x$ axis. This reduced system of 2D analogous Chern insulator is described by the reduced Bloch Hamiltonian $H_{k_z}(k_x,\theta)$, and the corresponding tight-binding chain Hamiltonian is given by
\begin{eqnarray}
&&\mathcal{\hat{H}}_{C} = \nonumber-\sum_{i_x}t_0\left(\hat{a}^{\dag}_{i_x+1,\uparrow}\hat{a}_{i_x,\uparrow}-\hat{a}^{\dag}_{i_x+1,\downarrow}\hat{a}_{i_x,\downarrow}\right)+\text{h.c.}\\ \nonumber&&~~~~~~~~~-\sum_{i_x}t_s\left(\hat{a}^{\dag}_{i_x,\uparrow}\hat{a}_{i_x+1,\downarrow}-\hat{a}^{\dag}_{i_x,\uparrow}\hat{a}_{i_x-1,\downarrow}\right)+\text{h.c.}\\ \nonumber&&~~~~~~~~~+\sum_{i_x}(M_z-2t_0\cos \theta)\left(\hat{a}^{\dag}_{i_x,\uparrow}\hat{a}_{i_x,\uparrow}-\hat{a}^{\dag}_{i_x,\downarrow}\hat{a}_{i_x,\downarrow}\right) \\ &&~~~~~~~~~+\sum_{i_x}2t_s\sin\theta\left(\hat{a}^{\dag}_{i_x,\uparrow}\hat{a}_{i_x,\downarrow}+\hat{a}^{\dag}_{i_x,\downarrow}\hat{a}_{i_x,\uparrow}\right).
\end{eqnarray}
\begin{figure}[tbph]
\includegraphics[width=8cm]{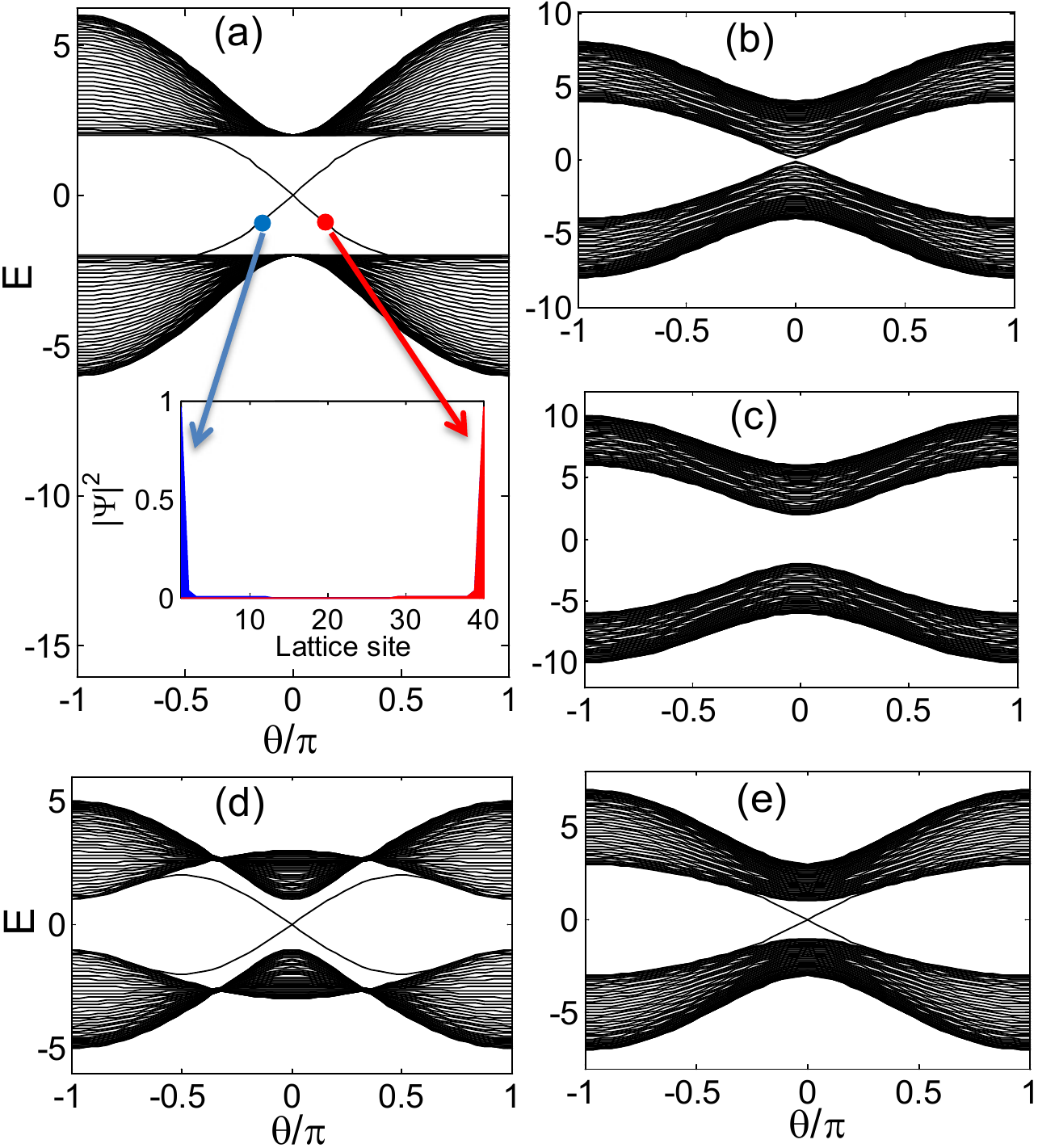}
\caption{(Color online) Energy spectrum of the reduced tight-binding chain of
analogous Chern insulator with lattice sites $L_x=40$ under open
boundary conditions. (a) non-trivial case for $k_z=0$ and $m_z=4$;
(b) critical case for $k_z=0.5\pi$ and $m_z=4$; (c) trivial case
for $k_z=\pi$ and $m_z=4$;  (d) non-trivial case for $k_z=0$ and
$m_z=3$; (e) non-trivial case for $k_z=0.5\pi$ and $m_z=3$. The
insert figure in (a) shows the density distribution of two typical
edge modes. Other parameters in (a-e) are $t_0=1$ as the energy
unit and $t_s=1$.}
\end{figure}

In Fig. 2, we numerically calculate the energy spectrum of the
reduced chain with length $L_x=40$ under open boundary
conditions for different parameters. From Figs. 2(a-c), we can
see the variation of the energy spectrum by changing the parameter
$k_z$ for fixed $m_z=4t_0$. For $k_z=0$ in Fig. 2(a), the spectrum
contains two symmetric bands with an energy gap opened, which
accompanies some in-gap modes splitting into two branches. The two
branches of edge modes cross and connect the separated bands. It
indicates that the system with the lower band filled is a
topologically nontrivial insulator with Chern number $C_{k_z}=-1$.
As we continuously increase the parameter $k_z$, the band gap
closes at the critical value $k_z=k_z^c=0.5\pi$ [Fig. 2(b)] and
then reopens without the in-gap edges modes [Fig. 2(c)],
indicating the system in the topologically trivial case with Chern
number $C_{k_z}=0$. Such a $k_z$-parameter induced transition
between topologically distinct regimes in numerical simulations is
consist with the previous analytical calculations. In Figs. 2(d)
and 2(e), we show the spectrum of topologically nontrivial states
for $k_z=0$ and $k_z=0.5\pi$ with fixed $m_z=3t_0$, in which cases
the critical value is $k_z^c=2\pi/3$. Despite the shapes of the
spectra being somewhat different, the existence of continuous edge
modes connecting the lower and upper bands indicates that the
system is topologically nontrivial in these cases. The density
distribution (profile) of the edge modes in these topologically
nontrivial cases are similar to the two typical ones shown in the
insert figure in Fig. 2(a), and they will gradually spread into
the bulk when their energies are closer to the bulk bands.

\begin{figure}[tbph]
\includegraphics[width=8cm]{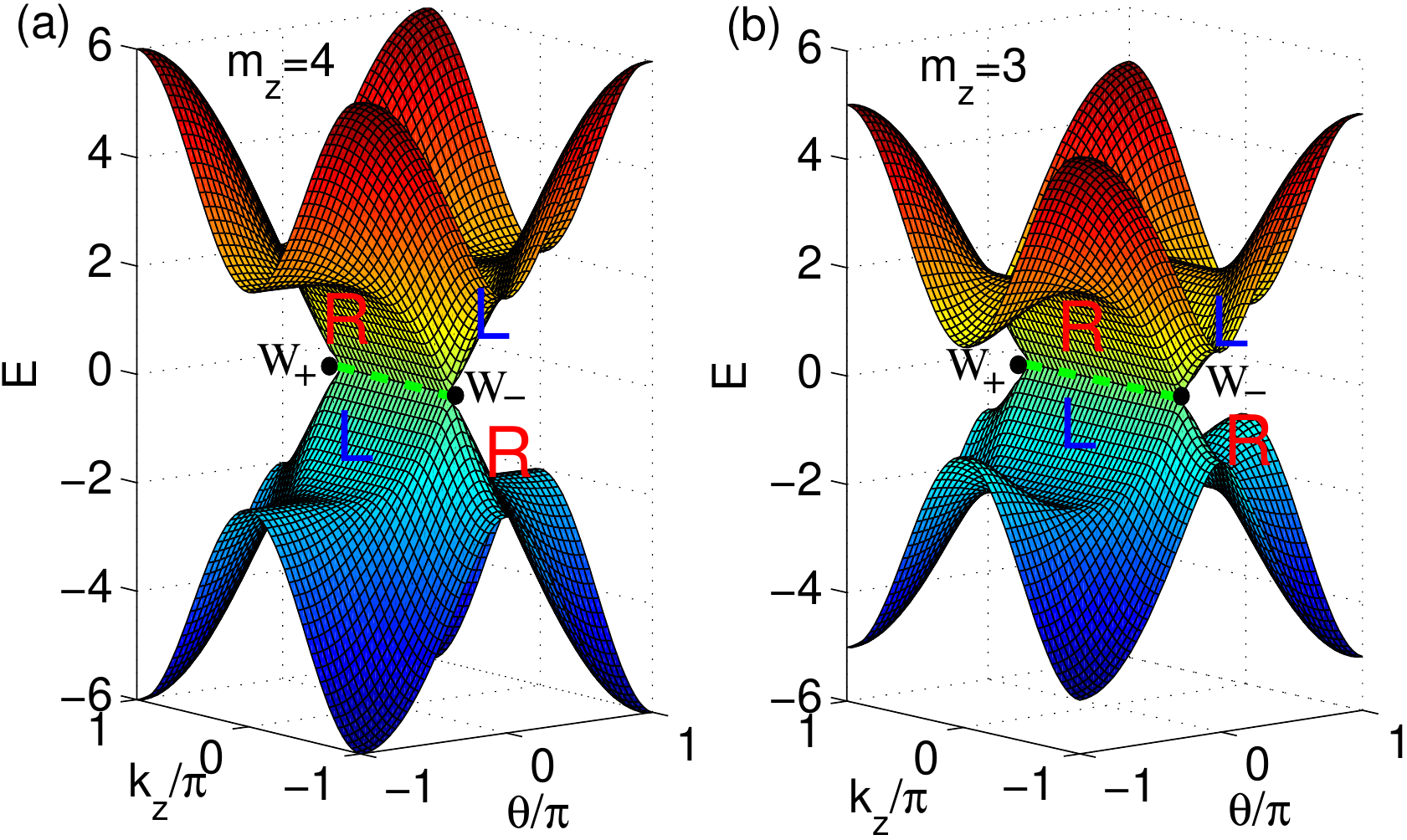}
\caption{(Color online) Energy spectrum of the edge states in the
mimicked WSM phase for (a) $m_z=4$ and (b) $m_z=3$.
The green dashed lines denote straight Fermi lines formed by gapless
zero-energy edge modes, which are analogous to Fermi arcs connecting the pair of the mimicked Weyl
points $\boldsymbol{W}_{\pm}$. They are located at
$(\theta,k_z)=(0,\pm\pi/2)$ in (a) and
$(\theta,k_z)=(0,\pm2\pi/3)$ in (b), respectively. The regimes of
left-edge and right-edge modes in the parameter space
denoted by blue 'L' and red 'R' are $\{E<0,\theta<0\}\bigcup\{E>0,\theta>0\}$ and $\{E>0,\theta<0\}\bigcup\{E<0,\theta>0\}$, respectively. The edge modes have similar profiles with those shown in Fig. 2(a). Other parameters in (a) and (b) are $t_0=1$ as
the energy unit and $t_s=1$.}
\end{figure}

In the context of 3D WSM phase, a pair of Weyl points
can be viewed as a monopole-antimonopole pair in the momentum space. As a consequence of these monopoles, there are
chiral Fermi surfaces with the energy $E=0$ connecting the Weyl
points on the surface not perpendicular to the $z$ axis, named as
Fermi arc \cite{Wan,Balents,Burkov,Xu,Delplace,Zhao}. In this 2D
system with an artificial dimension, in a similar way by assuming
that $\theta$ and $k_z$ are good quantum numbers, we can consider
edge modes along the $x$ axis in the mimicked WSM
phase. The system in this case returns to the reduced chain
described by Hamiltonian (15) and we find that analogous Fermi-arc
zero modes emerge by numerically calculating the energy spectrum
of the edges modes, as shown in Fig. 3. In this case, the pair of the analogous
Weyl points are connected with straight Fermi lines consisting of gapless zero-energy edge modes, which are analogous to
Fermi arcs in the $\theta-k_z$ parameter space (denote by green dashed lines). The corresponding regimes for left-edge and right-edge modes are
also shown in Fig. 3(a) and 3(b). The profiles of these edge modes are
similar with those shown in Fig. 2(a), and the edge modes closer
to the Weyl points and the bulk bands spread more into the bulk
than those in the center of the parameter regime.

\section{experimental detection of mimicked WSM phase}

So far, we have introduced the system for simulating the WSM physics and explored the relevant topological
properties. In this section, we propose practical methods for
their experimental detection in this cold atom system. We
first show that the mimicked Weyl points can be probed by
measuring the atomic Zener tunneling to the excited band after a
Bloch oscillation, and then propose a feasible technique to obtain
the $k_z$-dependent Chern number directly from the center shift of
the hybrid Wannier functions.

\subsection{Detection of the Weyl points}

The band touching points can be monitored from the atomic fraction
tunnelling to the excited band in Bloch oscillations, as recently
experimentally demonstrated to probe the Dirac points and the
topological phase transition in a honeycomb optical lattice
\cite{Tarruell,Lim}. This Bloch-Zener-oscillation technique can be
extended to detect the simulated Weyl points in 3D optical
lattices \cite{He} and in our system. Let us consider a 2D cloud
of noninteracting fermions prepared in the ground band in this
system and an external constant force $F$ is applied along the
$\eta$ ($\eta=x,z$) direction, which pushes the atoms moving along
the $k_{\eta}$ direction. For fixed parameters $\theta$ and $m_z$,
one can observe the quasi-momentum distribution of the transfer
fraction in the excited band from the time-of-flight measurements
after a Bloch cycle \cite{Tarruell,Lim}. One can perform such a
measurement for varying parameters, and finally the mimicked Weyl
points in the extended 3D Brillouin zone can be verified.

Without loss of generality, here we consider the case of a singe
pair of Weyl points described by the low-energy effective
Hamiltonian (12).  In this case, the transfer fraction
$\xi_x(k_z)$ [$\xi_z(k_x)$] along the $k_x$ ($k_z$) direction for
$k_z$-trajectory ($k_x$-trajectory) can be written as
\cite{Tarruell,Lim}
\begin{eqnarray}
&&\xi_x(k_z)=P_{LZ}^x(k_z), \\
&&\xi_z(k_x)=2P_{LZ}^z(k_x)\Big[1-P_{LZ}^z(k_x)\Big],
\end{eqnarray}
where $P_{LZ}^x=e^{-\pi\Delta_x^2(k_z)/4v_{x}F}$ and
$P_{LZ}^z=e^{-\pi\Delta_z^2(k_x)/4v_{z}F}$ are the Landau-Zener
transition  probabilities, with $\Delta_x=2E_{+}(k_x=0,k_z)$ and
$\Delta_z=2E_{+}(k_x,k_z=k_z^c)$ denoting the energy gaps for the
Landau-Zener events along the $k_{\eta}$ direction. For
simplicity, here we treat trajectories for different $k_{\eta}$ as
independent by neglecting the harmonic trap and assuming the two
tunneling events along the $k_z$ direction incoherent. Such a
simplified model was shown to be sufficient for comparison with
the experiments under realistic conditions \cite{Tarruell,Lim}.

\begin{figure}[tbph]
\includegraphics[width=8.6cm]{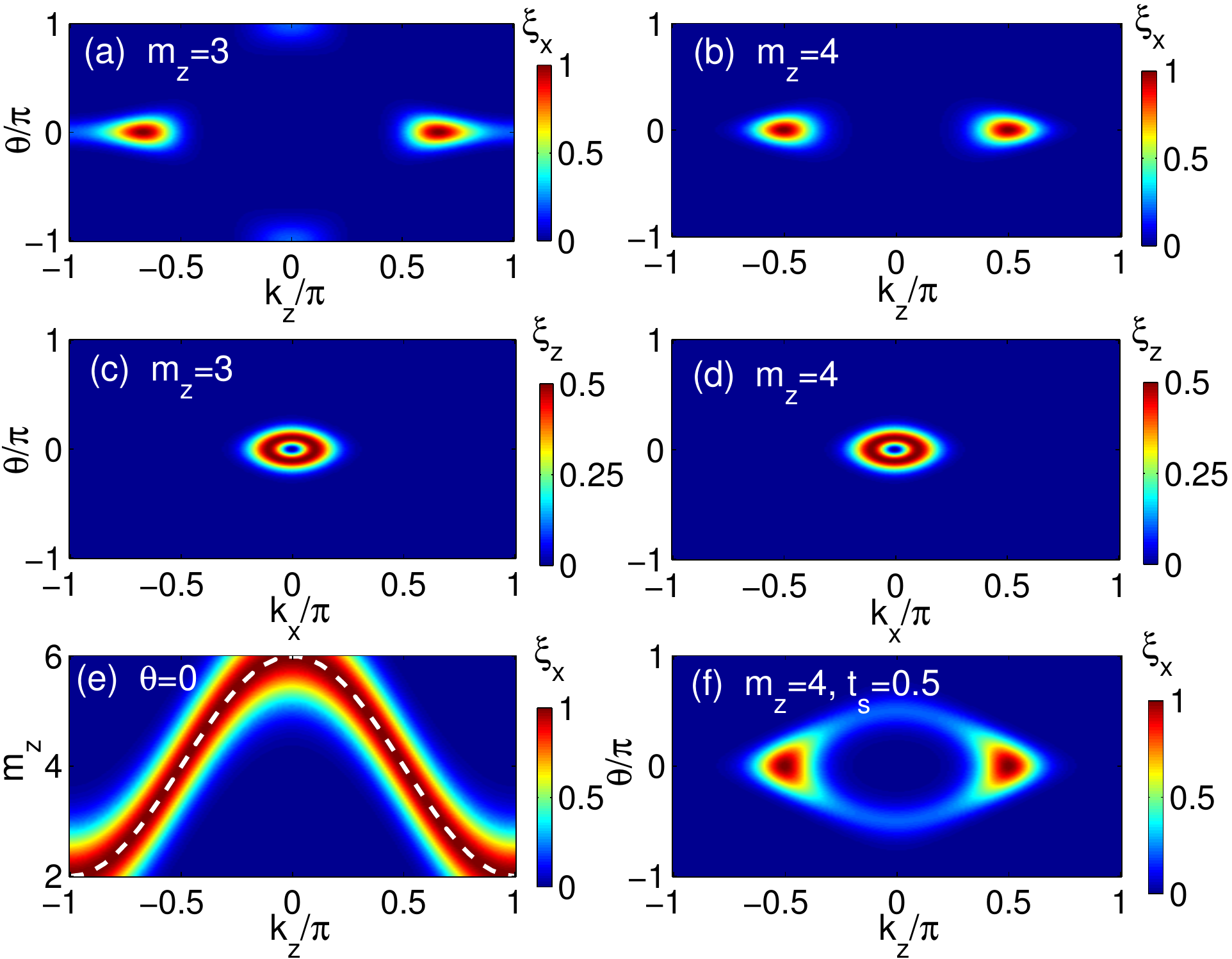}
\caption{(Color online) The transfer fractions for tunable parameters. The quasi-momentum distribution $\xi_{x}(k_z)$ for varying parameter $\theta$ with (a) $m_z=3$ and (b) $m_z=4$; the quasi-momentum distribution $\xi_{z}(k_x)$ for varying parameter $\theta$ with (c) $m_z=3$ and (d) $m_z=4$; (e) $\xi_{x}(k_z)$ for different parameter $m_z$ with $\theta=0$; (f) $\xi_{x}(k_z)$ for different parameter $\theta$ with $m_z$ and $t_s=0.5$. The white dashed line in (e) shows the expected $k_z$ positions of the mimicked Weyl points $\boldsymbol{W}_{\pm}$ for varying $m_z$, which are well agree with the maximum transfer positions of $\xi_x(k_z)$. The maximum $\xi_x(k_z,\theta)$ in (a) and (b) also correspond to the $k_z$ and $\theta$ positions, as shown in Fig. 3. The position of maximum dip inside the ring profile of $\xi_z(k_x,\theta)$ in (c) and (d) indicates $k_x=\theta=0$ for the Weyl points. Other parameters in (a-f) are $t_0=1$ as the energy unit, $F=1$, and $t_s=1$ except that $t_s=0.5$ in (f).}
\end{figure}

In Fig. 4, we calculate the transfer fractions $\xi_{x,z}$ using Eqs. (16) and (17) for some typical parameters. The quasi-momentum distributions $\xi_{x}(k_z)$ for varying parameter $\theta$ in Figs. 4(a) (with $m_z=3$) and 4(b) (with $m_z=4$) show two maximum transfer positions in the whole $k_z-\theta$ space, which correspond to the $k_z$ and $\theta$ positions of the paired Weyl points as excepted. These peaks of the maximum transfer factions are sharp since the transition probability in a single Landau-Zener event increases exponentially as the energy gap decreases. In other words, the emergence of the mimicked Weyl points, at which the band gap closes, is accompanied by a dramatic increase in the transfer fraction $\xi_{x}(k_z,\theta)$. In contract, there are two subsequent Landau-Zener transitions along the $k_z$ direction. This results in the distribution of the final transfer $\xi_{z}(k_x,\theta)$ exhibiting the ring-type profile of the maximum value $\xi_{z}=0.5$, as shown in Figs. 4(c) and 4(d). The position of maximum dip with the value $\xi_z\approx0$ inside the ring profile of $\xi_z(k_x,\theta)$ indicates $k_x=\theta=0$ for the mimicked Weyl points in these cases. We also plot the distribution $\xi_{x}(k_z,m_z)$ for fixed $\theta=0$ in Fig. 4(e), where the maximum transfer positions of $\xi_x(k_z)$ correspond well to the expected $k_z$ positions of the mimicked Weyl points as plotted by the dashed line. The $\xi_{x}(k_z,\theta)$ distributions for different $t_s$ will be modified, but the peaks of the transfer fractions remains, with an example shown in Fig. 4(f). Therefore, combining with the measurements of transfer distributions $\xi_{x}(k_z,\theta)$ and $\xi_{z}(k_x,\theta)$ from Bolch-Zener transitions along the $k_{x}$ and $k_{z}$ direction, one can resolve the mimicked Weyl points in the $\boldsymbol{\tilde{k}}$ space for different $m_z$. This detection method is robust against fluctuations of the hopping strengths, which are engineered by laser beams in this cold atom system. In addition, since the creation or annihilation of the analogous Weyl points by external parameters (such as $m_z$) will accompany topological phase transition, then the critical point of the transition can be identified by using this method.

\subsection{Detection of the $k_z$-dependent Chern number}

In solid-state materials, the Chern number value can be revealed from a routine measurement of the Hall conductance, however, the detection method is usually quite different in cold atom systems \cite{Price,Liu2013,Jotzu,Bloch2015,Demler,Bloch2013,Duca,Alba,Deng}. Since the Chern number comes from the integral of Berry curvature over the first Brillouin zone, one way to reveal the topological order is to measure the Berry curvature. It was theoretically proposed and experimentally performed in optical lattices to map the Berry curvature from the transverse drift induced by the anomalous velocity in Bloch oscillations \cite{Price,Liu2013,Jotzu,Bloch2015} and from an atomic interferometry in momentum space \cite{Demler,Bloch2013,Duca}. Besides, it was also proposed to measure the Berry curvature from atomic momentum distributions for different spin and spin-mixing components \cite{Alba,Deng}. These methods for measuring the Berry curvature could be potentially extended to our proposed cold atom system, as one can perform similar measurements for a fixed value of the parameter $\theta$.

Below we propose a practical method to directly probe the $k_z$-dependent Chern number given by Eq. (14) in our system, based on a generalization of topological pumping in optical lattices \cite{Thouless,Wanglei,Pumping2015}. Remind that in order to obtain the topological invariant, we treat our 2D system as a collection of tight-binding chains along the $x$ axis described by Hamiltonian (15), which can realize the analogous Chern insulators defined in the $k_x-\theta$ space as different slices of parameter $k_z$. The polarization of this 1D insulator can be expressed by means of the centers of the hybrid Wannier functions (HWFs) \cite{Marzari}, which are localized in the $x$ axis retaining Bloch character in the $\theta$ artificial dimension in our case. This polarization depends on the parameter $k_z$ and is a function of $\theta$, which acts as an external parameter under which the polarization changes. When $\theta$ is adiabatically changed by $2\pi$, the change in polarization, i.e., the shift of the HWF center, is proportional to the Chern number \cite{Vanderbilt,Marzari}. This is a manifestation of topological pumping \cite{Thouless}, with $\theta$ being the adiabatic pumping parameter. In our system, the HWF center for a tight-binding chain in Hamiltonian (15) is given by \cite{Wanglei}
\begin{eqnarray}
\langle n_x(\theta,k_z) \rangle = \frac{\sum_{i_x}i_x\rho(i_x,\theta,k_z)}{\sum_{i_x}\rho(i_x,\theta,k_z)},
\end{eqnarray}
where $i_x$ is the lattice-site index, and $\rho(i_x,\theta,k_z)$ is the density of the HWF and denotes the atomic densities resolved along the $x$ direction as a function of $\theta$ and $k_z$. Here the hybrid density can be written as
\begin{eqnarray}
\rho(i_x,\theta,k_z)=\sum_{\text{occupied states}}|i_x,\theta,k_z\rangle\langle i_x,\theta,k_z|,
\end{eqnarray}
where $|i_x,\theta,k_z\rangle$ is the hybrid eigenstate of the system. Experimentally, $\rho(i_x,\theta,k_z)$ can be measured by the hybrid
time-of-flight images \cite{Wanglei}, referring to an {\sl in situ} measurement of the density distribution of the atomic cloud in the $x$ direction during free
expansion along the $z$ direction for a fixed $\theta$.

\begin{figure}[tbph]
\includegraphics[width=8.6cm]{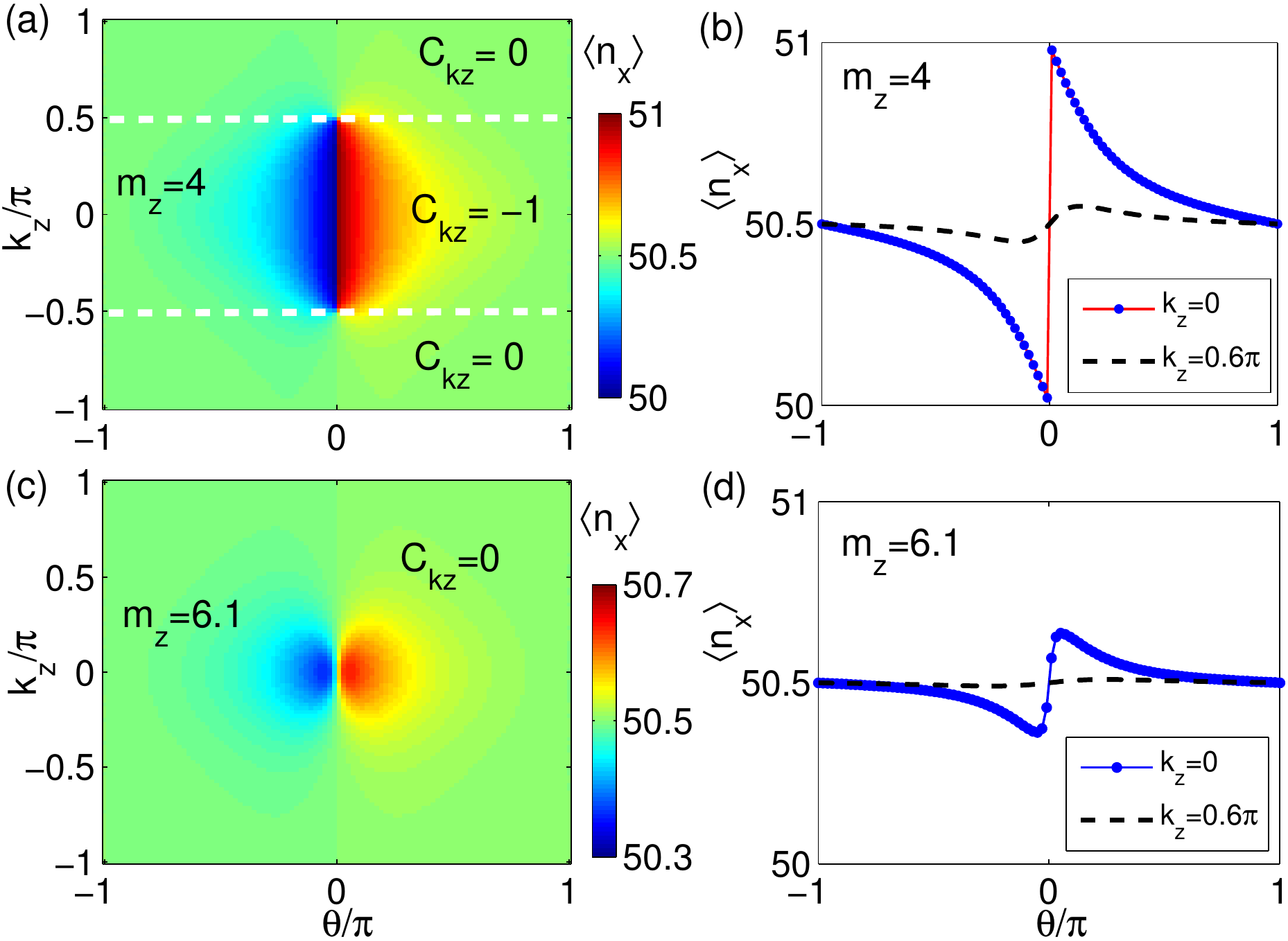}
\caption{(Color online) The HWF centers in a tight-binding chain of length $L_x=100$ at half-filling as a function of the adiabatic pumping parameter $\theta$ for different $k_z$. (a) The profile $\langle n_x(\theta,k_z)\rangle$ for the parameter $m_z=4$, where $\langle n_x(\theta)\rangle$ (do not) shows a jump of one unit cell for $k_z$ (outside) within the region $(-0.5\pi,0.5\pi)$, with typical examples shown in (b). (c) The profile for $m_z=6.1$ shows no jump of $\langle n_x(\theta)\rangle$ for all $k_z$, with typical examples shown in (d). The corresponding $k_z$-dependent Chern number $C_{k_z}$ is also plotted both in (a) and (c), with the white dashed lines denoting the critical value between the trivial ($C_{k_z}=0$) and nontrivial ($C_{k_z}=-1$) cases. Other parameters in (a-d) are $t_0=1$ as the energy unit and $t_s=1$.}
\end{figure}
\begin{figure}[tbph]
\includegraphics[width=8cm]{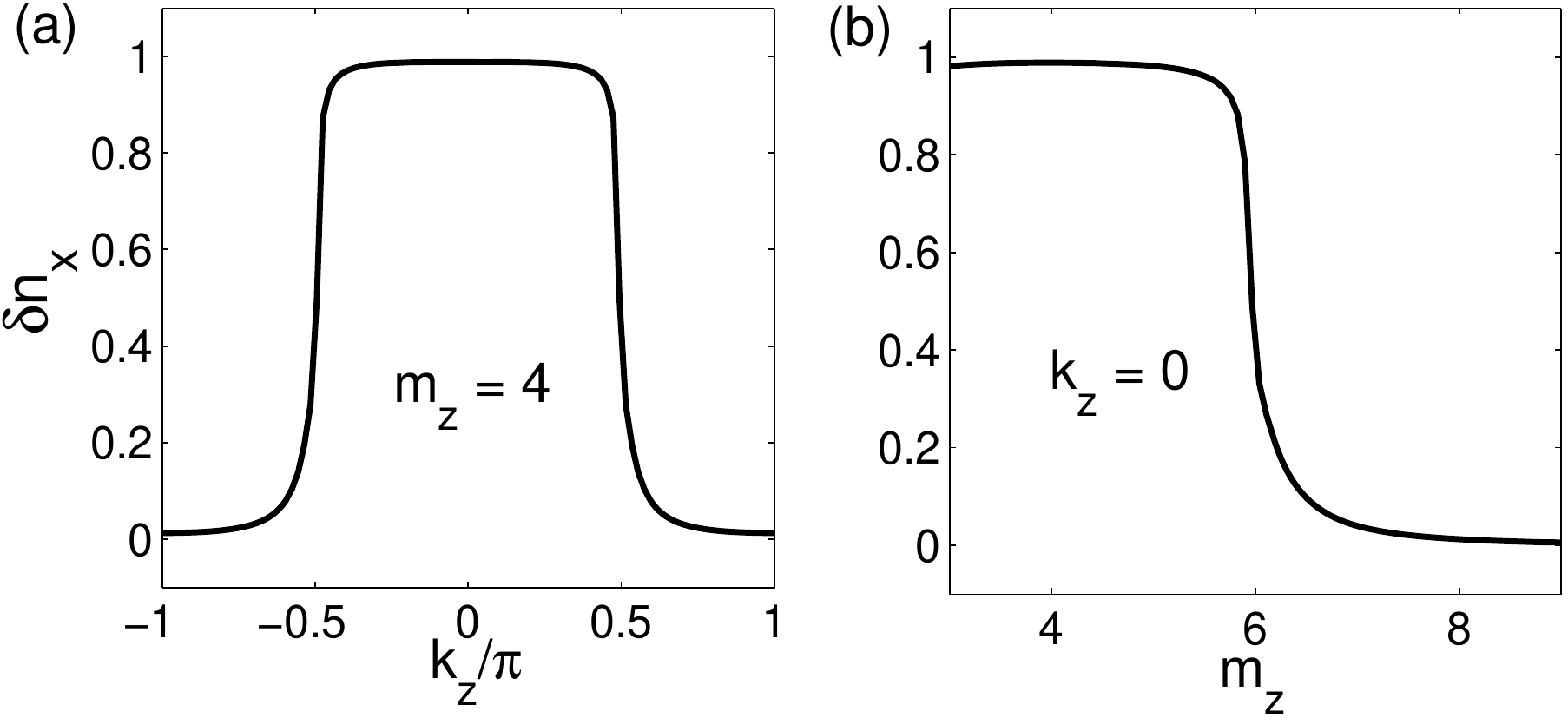}
\caption{ The maximum variation of the HWF-center shift $\delta n_x$. (a) $\delta n_x$ as a function of $k_z$ for $m_z=4$. (b) $\delta n_x$ as a function of $m_z$ for $k_z=0$. Other parameters in (a) and (b) are $L_x=100$, $t_0=1$ as the energy unit, and $t_s=1$.}
\end{figure}

We numerically calculate $\langle n_x(\theta,k_z)\rangle$ in a tight-binding chain of length $L_x=100$ at half-filling (assuming the Fermi energy $E_F=0$), with the results for typical parameter $m_z$ being shown in Fig. 5. For the case of $m_z=4$ in Figs. 5(a) and 5(b), the expected critical values between the trivial insulator with $C_{k_z}=0$ and Chern insulator with $C_{k_z}=-1$ are $k_z=\pm k_z^c=\pm0.5\pi$. As shown in Fig. 5(a), $\langle n_x(\theta)\rangle$ shows discontinuous jumps of one unit cell within the region $k_z\in(-0.5\pi,0.5\pi)$, which disappear outside this region. To be more clearly, we plot $\langle n_x(\theta)\rangle$ for $k_z=0$ and $k_z=0.6\pi$ as two examples in Fig. 5(b). The one-unit-cell jump driven by $\theta$ indicates that a single particle is pumped across the system \cite{Vanderbilt,Marzari}, as expected for $C_{k_z}=-1$. The results for $m_z=6.1$ show no jump of $\langle n_x(\theta)\rangle$ for all $k_z$ in Figs. 5(c) and 5(d), which is consistent with the expected $C_{k_z}=0$ for $m_z>6$ in this case. This establishes a direct and clear connection between the shift of the hybrid density center and the topological invariant. Thus one can directly extract the $k_z$-dependent Chern number from the hybrid time-of-flight images in this cold atom system \cite{Wanglei}. The proposed detection method based on topological pumping is not affected by a weak harmonic trap as it does not require the presence of the sharp
edge states \cite{Wanglei,Pumping2015}.

In the HWF-center shifts $\langle n_x(\theta)\rangle$ as shown in Fig. 5, the one-unit jump for topologically nontrivial cases and the continuous variation for trivial cases may be hard to distinct in realistic experiments. It will be more practical to extract the topological invariant from the maximum variation of $\langle n_x(\theta)\rangle$ during adiabatically varying $\theta$
\begin{eqnarray}
\delta n_x \equiv \max\{\langle n_x(\theta)\rangle\}-\min\{\langle n_x(\theta)\rangle\}.
\end{eqnarray}
In Fig. 6. we numerically calculate $\delta n_x$ as a function of $k_z$ ($m_z$) with fixed $m_z=4$ ($k_z=0$) for a tight-binding chain. One can find that $\delta n_x\simeq1$ in the parameter regimes where the system is topologically nontrivial, i.e. $k_z\in(-0.5\pi,0.5\pi)$ in Fig. 6(a) and $m_z<6$ in Fig. 6(b), while $\delta n_x\simeq0$ outside these regimes, corresponding to the trivial cases. Therefore, as long as the system is not close to the critical points, such as $k_z=\pm0.5\pi$ in Fig. 6(a) and $m_z=6$ in Fig. 6(b), one can extract the topological invariant in such a measurement as $\delta n_x=|C_{k_z}|$. This simple relationship does not well survive near the topological transition points, where the topological invariant will be more difficult to determine in experiments. It is interesting to note that the proposed measurement of Chern number is straightforward and the quantized pumping approach in the topologically nontrivial (trivial) cases, corresponding to $\delta n_x=1$ ($\delta n_x=0$), is robust against weak perturbations under realistic conditions \cite{Thouless,Wanglei}.

Finally, we note that the probe of topological edge states in a cold atom system, such as those shown
in Fig. 2 and Fig. 3, are also cumbersome. These edge states will
generally be washed out by the smooth harmonic potential and thus
one may not be able to distinguish them from the bulk states. This problem
could be circumvented potentially by using a steep confining
potential and imaging the edge states from Bragg signals by
means of specifically tuned Raman transitions \cite{Goldman2012}
or from their dynamics after suddenly removing the potential
\cite{Goldman2013}. If the analogous Fermi line states in Fig. 3 can be created, one may further
explore the induced topological current in the presence of an additional gauge field acting on the two Weyl points, which is used to break their central symmetry. The simplest example of such a gauge field can be generated by a constant Peierls phase for hopping along the $z$ axis induced by the laser-atom coupling.

\section{conclusions}

In summary, we have proposed an experimental scheme to simulate
and explore 3D topological WSMs with cold atoms in a 2D
square optical lattice subjected to an artificial dimension from
an external cyclical parameter. We have shown that this system is
able to describe the essential physics of WSMs with tunable
Weyl points, and have investigated the relevant topological
properties by calculating the Chern number and the gapless edge
states. Furthermore, we have proposed practical methods for the
experimental detection of the mimicked Weyl points and the
characteristic topological invariant in this cold atom system.
Considering that all the ingredients to implement our scheme in
the optical lattice have been achieved in the recent experiments,
it is anticipated that the presented proposal will be tested in an
experiment in the near future. Our proposed system would provide a
promising platform for elaborating the intrinsic exotic physics of
WSMs that are elusive in nature.

\section{Acknowledgements}

We thank T. Mao, Y. X. Zhao, J. Zhang, F. Mei, and Z.-Y. Xue for helpful discussions. This work was supported by the GRF (Grants No.
HKU7045/13P and HKU173051/14P) and the CRF (Grant No. HKU8/11G) of Hong Kong, the SKPBR of China (Grant No. 2011CB922104), the NSFC (Grants No. 11125417 and 11474153), and the PCSIRT (Grant No. IRT1243).

\end{document}